\documentclass{iau}
\usepackage{graphicx, color}

\usepackage{multicol}
\usepackage{natbib}

\title[Radio and $\gamma$-ray NLS1s in the spotlight] 
{Radio and $\gamma$-ray loud narrow-line Seyfert 1 galaxies in the spotlight}

\author[V. Karamanavis et al.]
{
V. Karamanavis$^{1,\star}$, 
 E. Angelakis$^1$,
 S. Komossa$^2$,
 I. Myserlis$^1$,\\
 D. Blinov$^3$,
 \and
 J. A. Zensus$^1$
 }

\affiliation{
$^1$
Max-Planck-Institut f\"ur Radioastronomie, Auf dem H\"ugel 69, 
D-53121 Bonn, Germany \\ email: {\tt vkaramanavis@mpifr.de} \\[\affilskip]
$^2$QianNan Normal University for Nationalities, Longshan Street,
Duyun, Guizhou, China\\[\affilskip]
$^3$Department of Physics and Institute of Theoretical and Computational 
Physics,\\ University of Crete, 71003 Heraklion, Greece
}

\pubyear{2016}
\jname{New Frontiers in Black Hole Astrophysics} 
\begin{document}

\maketitle

\begin{abstract}
Narrow-line Seyfert 1 (NLS1) galaxies provide us with unique insights into the 
drivers of AGN activity under extreme conditions. Given their low black hole 
(BH) masses and near-Eddington accretion rates, they represent a class of 
galaxies with rapidly growing supermassive BHs in the local universe. 
Here, we present the results from our 
multi-frequency radio 
monitoring 
of a sample of $\gamma$-ray loud NLS1 galaxies ($\gamma$NLS1s), 
including systems discovered only recently, and featuring both the 
nearest and the most distant $\gamma$NLS1s known to date. We also 
present high-resolution radio imaging of 
1H 0323+342, which is 
remarkable for its spiral or ring-like host.
Finally, we present 
new radio data of the candidate $\gamma$-emitting NLS1 galaxy RX J2314.9+2243, 
characterized by a very steep radio spectrum, unlike 
other $\gamma$NLS1s.
\end{abstract}

\firstsection 
\section{Introduction}
Narrow-line Seyfert 1 galaxies are a particular class of active galactic nuclei 
(AGN) that show:
(i) small width of their broad optical emission lines, i.e. 
FWHM(H$_{\beta}) \leq 2000$\,km\,s$^{-1}$ \citep{1985ApJ..297...166}, pointing 
to a low black hole (BH) mass for these systems (${\sim} 
10^{6}{-}10^{8}$\,M$_{\odot}$),
(ii) super-strong Fe\,II emission complexes,
(iii) rapid X-ray variability, hinting that the broad-line region (BLR) and 
accretion disk are directly visible,
(iv) near-Eddington accretion rates with ratios $L/L_{\rm{Edd}}$ 
between 0.1 and 1 \citep{1992ApJS...80..109B},
(v) super-soft X-ray spectra, and
(vi) a multitude of intriguing multi-wavelength properties 
\citep[see review by][]{2008RMxAC..32...86K}.

There exists also an interesting small fraction of their population being 
radio-loud, featuring relativistic jets, and detected at $\gamma$ rays by 
\textit{Fermi} 
\citep[$\gamma$NLS1s;][]{2006AJ....132..531K,2009ApJ...707..727A,
2009ApJ...707L.142A}.
These few sources, the radio- and $\gamma$-ray-loud population, are exceptional 
because they show blazar-like observational attributes such as flat radio 
spectra, high brightness temperatures (T$_{\rm b} \sim 
10^{10}{-}10^{14}$\,K 
), Doppler boosting, $\gamma$-ray emission and in few cases, 
one-sided 
jets \citep{2013MNRAS.436..191D,2015A&A...575A..55A,
2015PhDT.......232K,2016arXiv160803232F}. However, possessing non-blazar 
physical properties, especially almost two 
orders of magnitude lower BH masses and high accretion rates. In this context, 
they constitute a source of new insights, able to give us clues on the 
formation of extragalactic jets and their evolution under conditions of high 
accretion rate and in a regime that is not probed by classical blazars.
As possible explanations for their appearance, different scenarios have been 
proposed. Do they represent a young AGN population that rapidly grow their 
BHs? And how their orientation, with respect to the observer's line of sight, 
enters this challenging equation?

In the following, we present our radio monitoring of 
$\gamma$NLS1s using primarily the 100-m telescope at Effelsberg and the 
IRAM 30-m telescope, while focusing on latest results. Furthermore, we focus on 
two $\gamma$NLS1s which have been studied
in detail: (i) RX\,J2314.9+2243 with dedicated multi-wavelength monitoring, and 
(ii) 1H\,0323+342 with the help of high-resolution very-long-baseline 
interferometry (VLBI).

\section{Radio monitoring of NLS1 galaxies}

\begin{figure}
\begin{center}
 \includegraphics[width=0.49\textwidth]{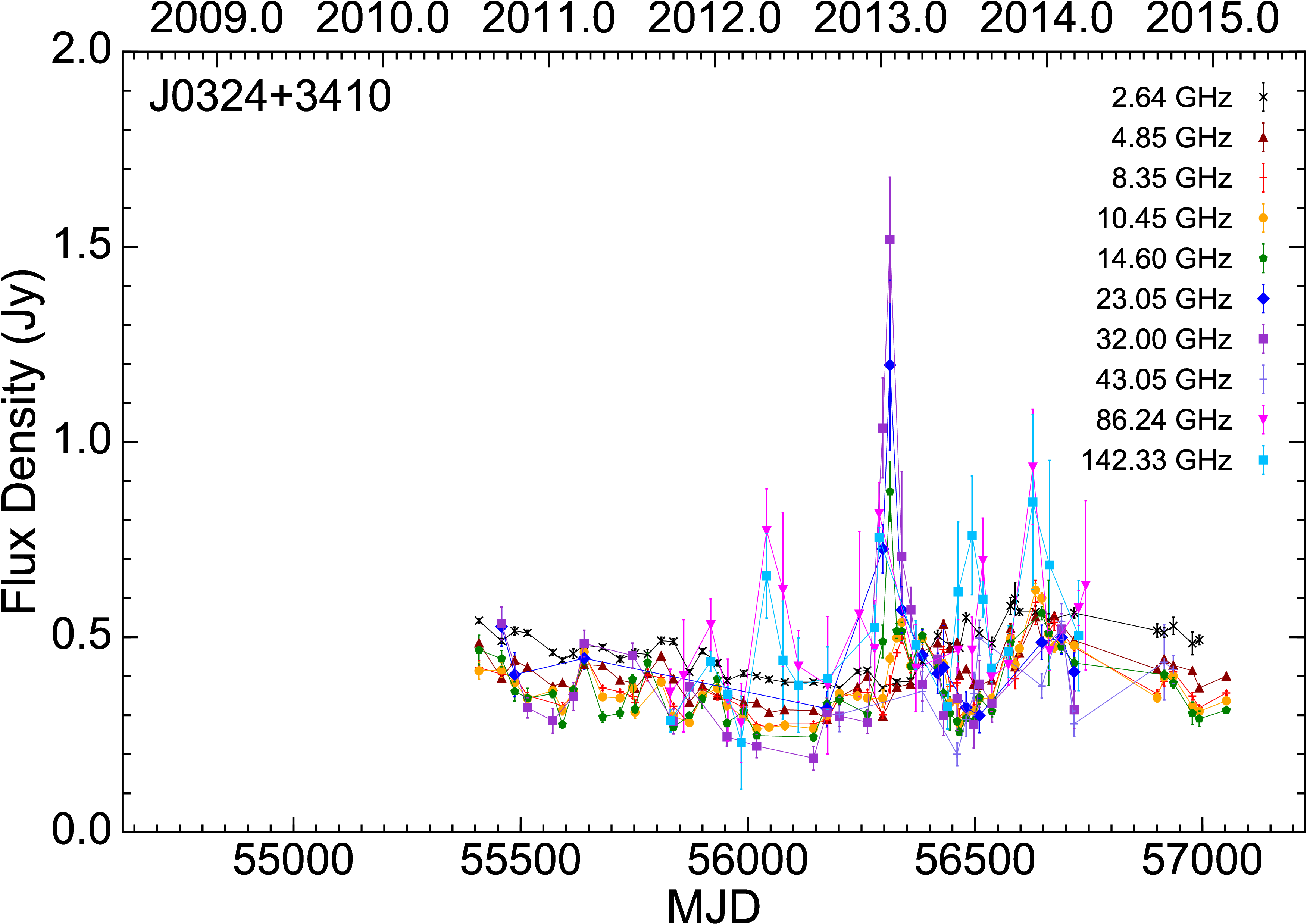}
 \hspace{1pt}
 \includegraphics[width=0.49\textwidth]{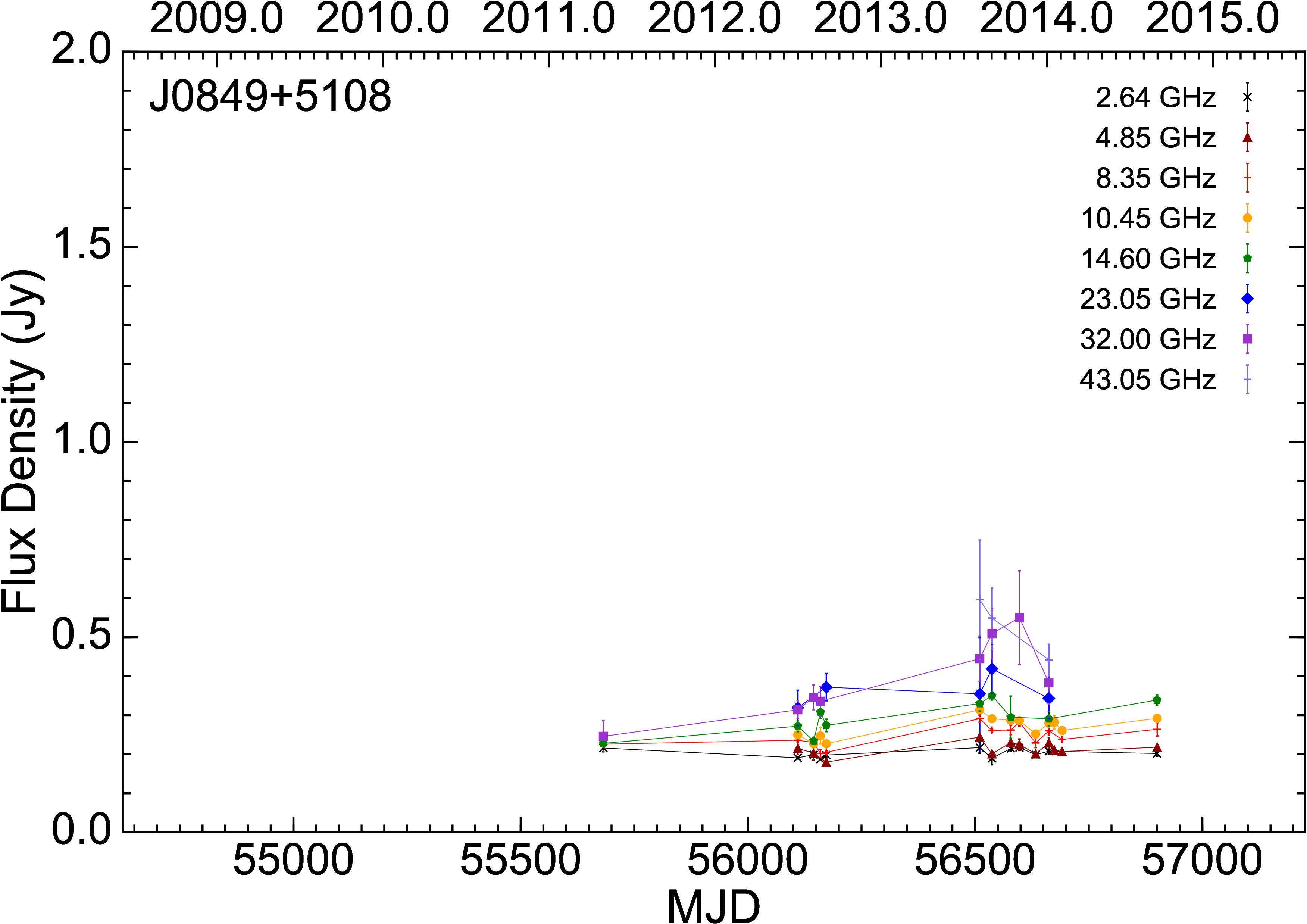}\\
 \vspace{4pt}
 \includegraphics[width=0.49\textwidth]{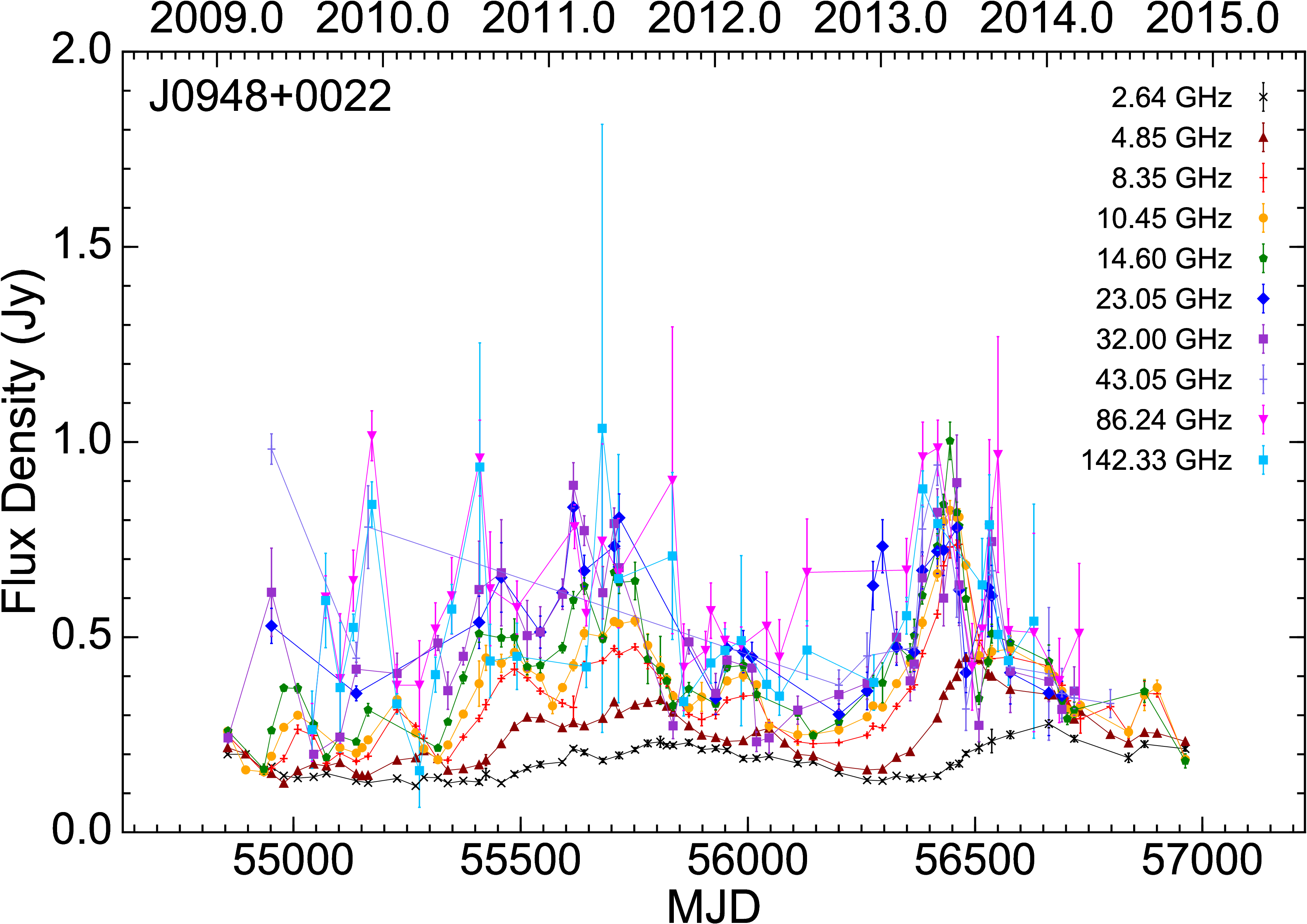}
 \hspace{1pt}
 \includegraphics[width=0.49\textwidth]{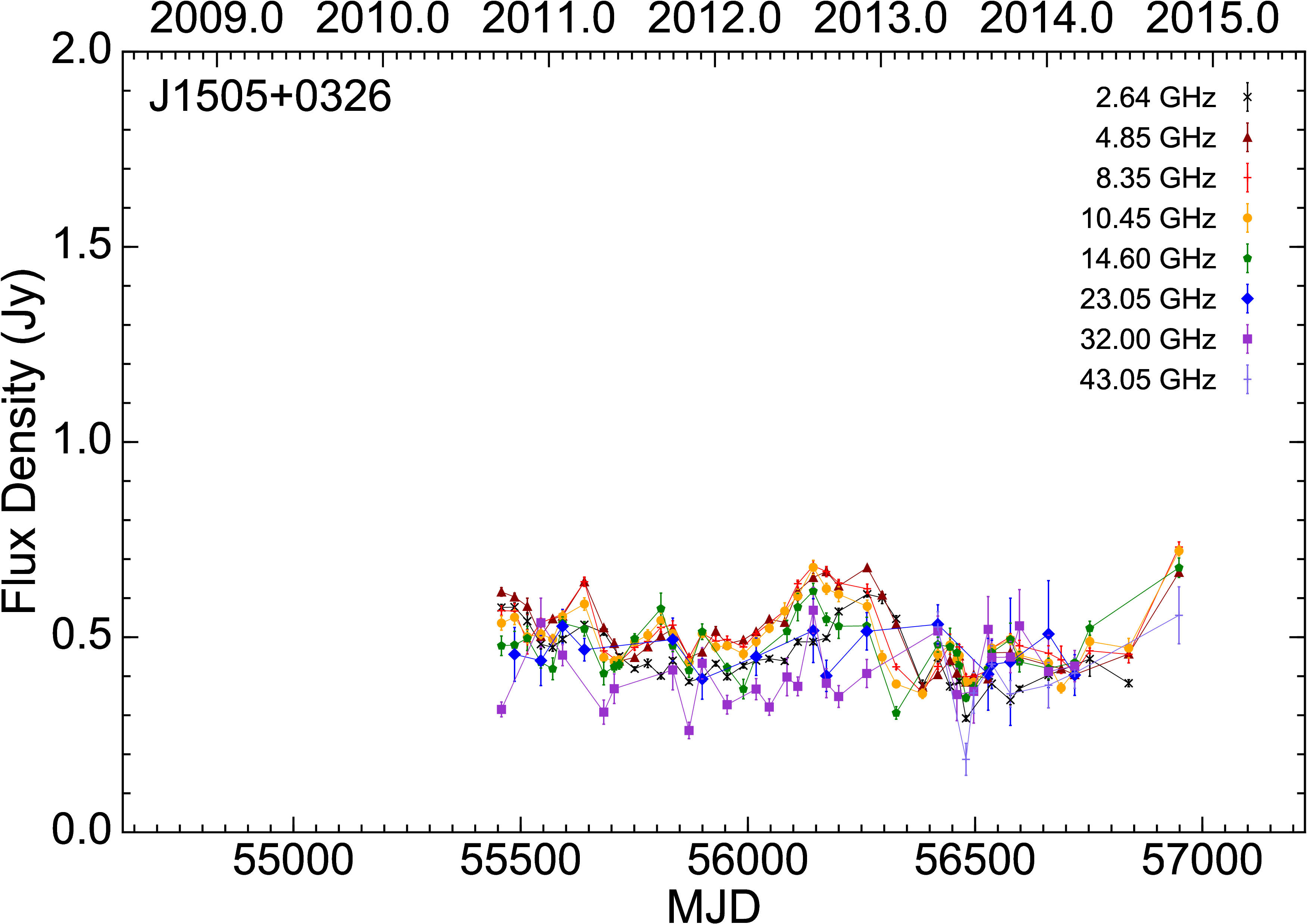}
 \caption{Updated radio light curves of $\gamma$NLS1s between 2.6 and 42\,GHz, 
until 2015 January. Where available, data at 86 and 142.33\,GHz are also shown.}
   \label{fig1}
\end{center}
\end{figure}

Our monthly monitoring of the jet emission from $\gamma$-ray-emitting NLS1 
galaxies constitutes the most comprehensive such program at cm and short mm 
wavelengths \citep{2015A&A...575A..55A}. Observations with the 
Effelsberg 100-m telescope cover eight bands between 2.64 and 43.05\,GHz. Until 
2014, the IRAM 30-m telescope provided coverage at 86.24 and 142.33\,GHz.
The data collected in more than five years is the longest data set with the 
widest frequency coverage available for these objects (see Fig. \ref{fig1}). 
It includes also radio polarisation monitoring at 2.64, 4.85, 8.35, and 
10.45\,GHz, with measurements of its linear and circular components and EVPAs. 
The observed sample comprises the following sources:
1H\,0323+342 (J0324+3410),
SBS\,0846+513 (J0849+5108),
PMN\,J0948+0022 (J0948+0022), and
PKS\,1502+036 (J1505+0326).
Recently, three more $\gamma$NLS1s were included, namely
FBQS\,J1644+2619,
SDSS\,J1222+0413, and
B3\,1441+476.

All sources show the typical blazar behaviour although 
with lower flux densities. $\gamma$NLS1s are variable and flare often, 
but the events are characterized by smaller amplitude and shorter duration. 
However, more energetic outbursts can be seen from time to time which dominate 
a source's light curve. Their spectra exhibit intense evolution 
with time, and steep, flat or even inverted spectra can be seen at different 
epochs. This evolution occurs faster than in blazars and can be ascribed to 
evolving shocks. The estimates of their jet powers are comparable with the 
output of the least energetic blazars. Furthermore, their moderate 
T$_{\rm b}$ and associated Doppler factors ($\delta \lesssim 10$), 
point to mildly relativistic jets as the driver of their phenomenology 
\citep{2015A&A...575A..55A}.

Most sources do not display detectable polarisation. However, 
1H\,0323+342 is $\sim$5.5\% linearly polarized, higher than what is 
typically observed in blazars \citep[e.g. 
$\sim$3.5--5\%][]{2015PhDT.......XXXM}. Its EVPA, at ${\sim}45^\circ$, lies 
perpendicular to the jet orientation, suggesting a magnetic field orientation 
almost parallel to the jet. PKS\,1502+036 appears to be polarized at 8.35 GHz 
($\sim$2--3\% with its EVPA at ${\sim}55^\circ$) during flaring periods. In the 
R band, $\gamma$NLS1s have been monitored with RoboPol since 2013 
\citep[e.g.][]{2016MNRAS.463.3365A}. Their mean linear polarisation ranges from 
less than 1\% to up to 20\%.

\section{The curious case of RX\,J2314.9+2243}

This is a mildly radio-loud (R = 10--20) NLS1 galaxy at a redshift z = 0.1692 
with a BH mass of 8$\times$10$^{7} $M$_{\odot}$ and a ratio 
$L/L_{\rm{Edd}} = 0.2$ \citep{2006AJ....132..531K}. What 
sets this source apart from the rest of 
the NLS1 population is its tentative detection at $\gamma$ rays by 
\textit{Fermi} (L. Foschini, priv. comm.) and its steep radio 
spectrum \citep[$\alpha = -0.76$;][]{2015A&A...574A.121K}.
The ongoing radio observations verify the 
steep spectrum with flux densities 
$\textrm{S}_{4\,\rm{GHz}}=10 \pm 2$\,mJy and
$\textrm{S}_{8\,\rm{GHz}}=6 \pm 1$\,mJy on two occasions in 2014 October.
To date, all $\gamma$-ray detected NLS1s have been associated with flat 
spectrum sources and RX\,J2314.9+2243 could be the first steep radio spectrum 
NLS1 to emit $\gamma$ rays.

\cite{2015A&A...574A.121K} address the emission processes 
shaping the spectral energy distribution (SED) of RX\,J2314.9+2243. 
The source is a luminous infrared emitter with an SED showing a broad 
emission hump between IR and UV bands, that steepens in the UV. An unusual 
feature of the source, compared to other NLS1s with steep X-ray spectra 
\citep[e.g.][]{1996A&A...305...53B,2004AJ....127.1799G,2006ApJS..166..128Z}, is 
the flatness of its X-ray spectrum, hinting at jet activity. In the UV on 
the other hand, the spectrum is very steep but there is no evidence of optical 
extinction beyond what is expected from our Galaxy, as would be expected by the 
dusty environment of luminous infrared galaxies (LIRGs). Therefore the IR to UV 
emission is most likely of non-thermal (i.e. synchrotron) origin arising from 
the operating jet.

Another striking feature of RX\,J2314.9+2243 is its very broad  
O[III] 5007\,\AA{} emission line component. This shows an unusually high 
kinematical blueshift of 1260 km\,s$^{-1}$ that could be due to the presence of 
an outflow powered by a radio jet feedback, in a face-on orientation with 
respect to our line of sight \citep{2015A&A...574A.121K}.

\section{Direct imaging of 1H\,0323+342 at 15\,GHz}

1H\,0323+342 is the most nearby radio-loud (R $\sim 50$) $\gamma$-ray 
emitting NLS1 at z $=0.0629$ \citep{2007ApJ...658L..13Z}. It exhibits a 
high Eddington luminosity ratio ($L/L_{\rm{Edd}} \sim  0.1$) with a low 
BH mass of ${\sim} 10^{7}$\,M$_{\odot}$ confirmed by several independent 
estimates \citep{2007ApJ...658L..13Z,2015AJ....150...23Y, 
2016ApJ...824..149W,2016arXiv160908002L}. 1H\,0323+342 is highly 
variable at radio bands (cm to mm; see Fig. \ref{fig1}) and is also a special 
case because of the morphology of its host galaxy  which appears ring-like or 
as a one-armed spiral \citep{2007ApJ...658L..13Z,2008A&A...490..583A, 
2014ApJ...795...58L}.

We studied the parsec-scale structure and kinematics of the source employing 
high-resolution VLBI images at 15\,GHz obtained in eight epochs from 2010 
October until 2013 July\footnote{Data from the 
MOJAVE monitoring program \citep{2009AJ....137.3718L}.}. 
Additionally, with our single-dish observations we infer the T$_{\rm b}$ and 
$\delta$ for the source. This combination provides a good 
estimate for the viewing angle towards 1H\,0323+342 \citep[see][for first 
results]{2015PhDT.......232K}.

1H\,0323+342 shows a core--jet morphology characterized by the prominent 
core, a straight jet, and a total absence of emission for the receding side 
of the jet; i.e. no counter-jet is detected.
A stationary feature is seen close to the core while six components move along 
the relativistic jet. Five jet components show superluminal motion with 
velocities from $~1$\,c up to $~7$\,c 
\citep{2015PhDT.......232K,2016arXiv160803232F}.

The source shows fast variability seen with single-dish and VLBI 
monitoring alike. The flaring events are of low to mild amplitude, but more 
rapid compared to the typical long-term behavior of blazars. 1H\,0323+342 shows 
high brightness temperatures. While \cite{2015A&A...575A..55A} 
report a Doppler factor of 3.6 at 14.6\,GHz, and a higher one $(\delta = 4.3)$ 
at 2.64\,GHz, we deduce a more extreme value of T$_{\rm b} = 
5.7\times10^{12}$\,K and an associated Doppler factor of 5.2, from the most 
rapid and largest flux density variation. Combining the knowledge on the source 
kinematics (i.e. the speeds of jet components) with the Doppler factors inferred 
from the observed variability, allowed us, under the assumption of causality, 
to constrain the viewing angle towards the source. The range of plausible 
values is $\theta \leq 4^{\circ}$--$13^{\circ}$ 
\citep{2015PhDT.......232K,2016arXiv160803232F}. 
The smaller value of this range 
is consistent with viewing angles typical of 
blazars, but nevertheless a larger $\theta$ with the respect to 
the observer's line of sight cannot be excluded.

\section{Concluding remarks}

The remarkable characteristics of NLS1 galaxies separate them as 
prominent class of AGN and made them, in recent years, the focal point of 
intensive research. These systems accrete matter at a rate near the 
Eddington limit and rapidly grow the low-mass black holes driving them.
The study of the small population of radio-loud and gamma-ray detected NLS1s 
probes a parameter space not constrained by classical blazars and can provide 
a deeper understanding on the formation and evolution of powerful 
radio jets and the interplay with their complex environments.



\end{document}